\documentclass[twocolumn,showpacs,preprintnumbers,amsmath,amssymb,floatfix,prstab]{revtex4}

\usepackage{graphicx}
\usepackage{subfigure}
\usepackage{dcolumn}
\usepackage{bm}
\usepackage{ctable}
\usepackage{multirow}

\begin{document}
\title{Extending Item Response Theory to Online Homework}
\author{Gerd Kortemeyer}
\email{kortemey@msu.edu}
\affiliation{%
Lyman Briggs College and Department of Physics and Astronomy,
Michigan State University, East Lansing, MI 48824, USA
}%
\date{\today}
\begin{abstract}
Item Response Theory becomes an increasingly important tool when analyzing ``Big Data'' gathered from online educational venues. However, the mechanism was originally developed in traditional exam settings, and several of its assumptions are infringed upon when deployed in the online realm. For a large enrollment physics course for scientists and engineers, the study compares outcomes from IRT analyses of exam and homework data, and then proceeds to investigate the effects of each confounding factor introduced in the online realm. It is found that IRT yields the correct trends for learner ability and meaningful item parameters, yet overall agreement with exam data is moderate. It is also found that learner ability and item discrimination is over wide ranges robust with respect to model assumptions and introduced noise, less so than item difficulty.
\end{abstract}
\pacs{01.50.H-,01.40.G-,01.40.-d,01.50.Lc}

\maketitle

\section{Introduction}
Item Response Theory (IRT) offers the opportunity for data-driven development and evaluation of assessment items, such as homework, practice, concept inventory, and exam problems. The technique is increasingly used in Physics Education Research, for example to examine the validity of concept tests~\cite{lin09,cardamone11}, but also to evaluate online homework problems (e.g.~\cite{lee08}).

Originally, IRT was developed with a number of assumptions, based on standard exam settings:
\begin{itemize}
\item Learners receive no correctness feedback while taking the exam, and thus effectively have only one attempt to get a problem correct.
\item The setting is high-stakes, so learners will refrain from guessing if at all possible.
\item Learners receive no outside help.
\item The setting is controlled, so copying of solutions is minimized.
\item The identity of the learner is verified, so the person taking the exam is indeed who they claim to be.
\item The assessment is summative and designed to evaluate the ability of the learner.
\end{itemize}
All of these assumptions are likely violated in online homework situations:
\begin{itemize}
\item Most systems offer correctness feedback and allow for multiple attempts.
\item Learners are guessing answers even if minimal effort could result in correct solutions~\cite{kortemeyer09}.
\item Learners receive outside help from online forums, TAs, older students, etc, some of which is productive and some of which is not~\cite{kortemeyer05ana,kortemeyer07correl}.
\item Students are copying answers~\cite{palazzo10}, most likely from students who already successfully solved the problem.
\item Students are solving each other's homework problems.
\item The assessment is formative and designed to increase the ability of the learner.
\end{itemize}
It is thus not clear that IRT can be expanded into the online homework realm. The question is of increasing importance as larger and larger online courses are viewed as sources of ``Big Data,'' and analytics is seen as an essential component of the next generation publication of educational materials. How large is the error introduced by each of the confounding factors? In the online realm, is IRT still a meaningful measure of problem difficulty and discrimination?

But IRT is not only relevant for material development and evaluation. In large courses, IRT has the potential of early detection of students-at-risk. Ideally, low ability learners should be identified before summative assessment venues, so that there is still a chance for remediation and additional support. Thus, data from online homework could be the key to helping these learners --- but how reliable is IRT based on online formative assessment data for the determination of learner ability?

Section~\ref{sec:datamethod} presents the population and setting of the study, as well as the employed statistical techniques. Section~\ref{sec:problem} illustrates the problem by comparing an IRT analysis of the exam data with an analysis of homework data. Section~\ref{sec:attempts} investigates the effect of having multiple attempts in the online realm, while Section~\ref{sec:guessing} investigates guessing and copying behavior. Finally, Section~\ref{sec:conclusions} provides the conclusions of this study.

\section{The Data and Method}\label{sec:datamethod}
To investigate the influence of violating some of the basic assumptions of IRT, this study considers a large enrollment (256 student) physics course for scientists and engineers. This course recently increased the number of exams throughout the semester (12 exams and one final)~\cite{laverty12b}, which resulted in 184 exam items. The course also has 13 online homework assignments with a total of 401 problems, which are administered via LON-CAPA~\cite{kortemeyer08}.

Like many other assessment systems, LON-CAPA randomizes problems --- students get different versions of the same problem, e.g., different numbers, graphs, formulas, scenarios, or images~\cite{kortemeyer08}. Problems can be rendered differently for different settings, e.g., a problem that was written as a free-response numerical problem appears as a multiple choice problem on printed exams, where either the system or the author specify randomized wrong answer choices.

Both exams and homework had a mixture of conceptual and numerical questions. Exams were administered in a lecture hall using bubble sheets and multiple choice questions, where each student had a different randomized version of the problems in order to minimize copying. Also for homework, each student had a different version of the problems to at least avoid blind copying of answers. LON-CAPA was configured to give immediate correctness feedback and to allow for multiple attempts (in case of multiple choice questions usually two or three attempts, in case of free-response numerical questions 99 attempts (i.e., essentially unlimited)). No grade penalty was given for using multiple attempts.

Unless noted otherwise, for IRT calculations, this study is using the widely accepted two parameter logistic (2PL) model~\cite{birnbaum}. Items are modeled using two parameters that correspond to difficulty and discrimination, and learners are modeled using an ability estimate. Calculations are performed using the Latent Trait Model ({\tt ltm}) package~\cite{ltm} within the {\tt R} statistical software system~\cite{rpackage}.

\section{The Problem}\label{sec:problem}
\subsection{Item Characteristics}
\begin{figure*}
\begin{center}
\includegraphics[width=0.48\textwidth]{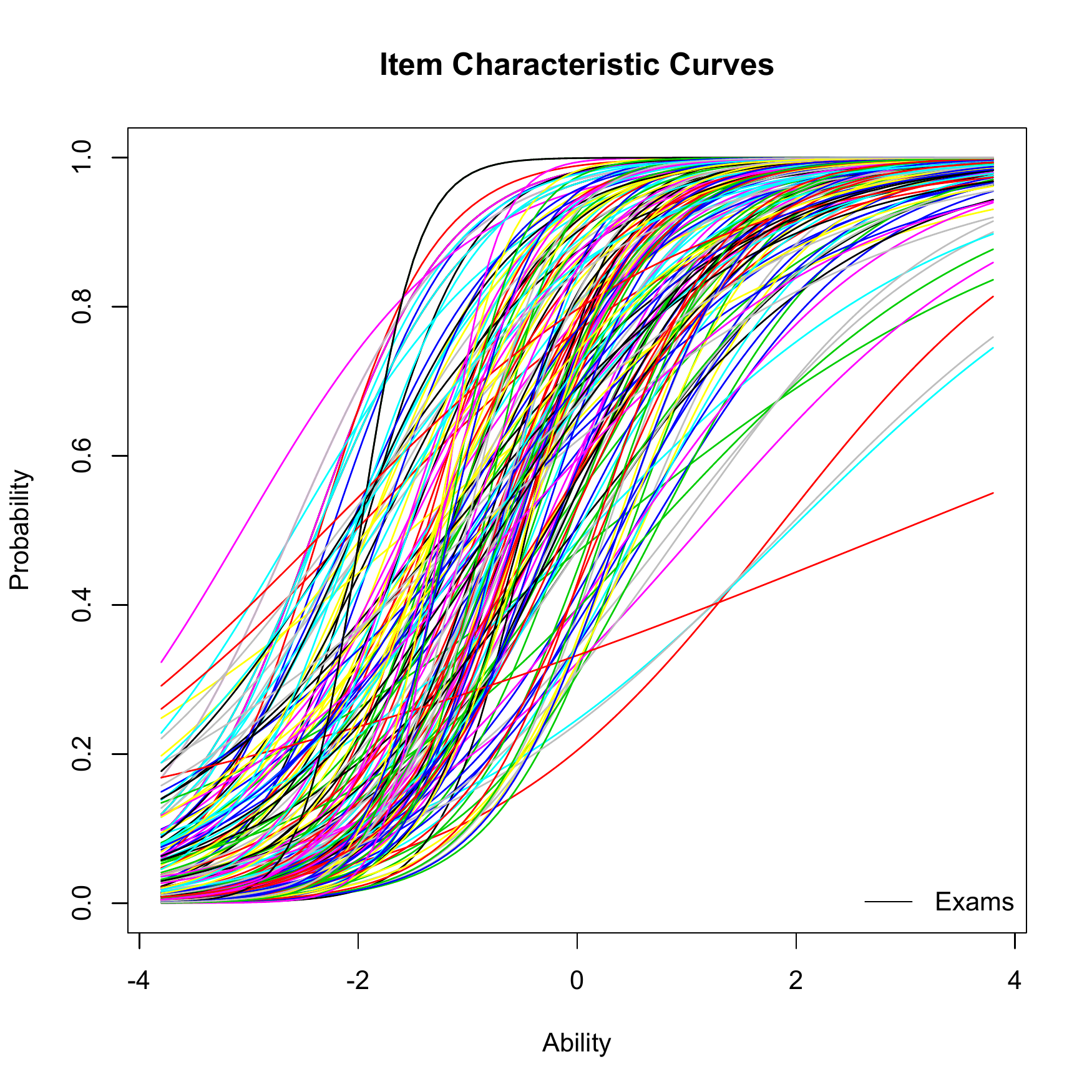}
\includegraphics[width=0.48\textwidth]{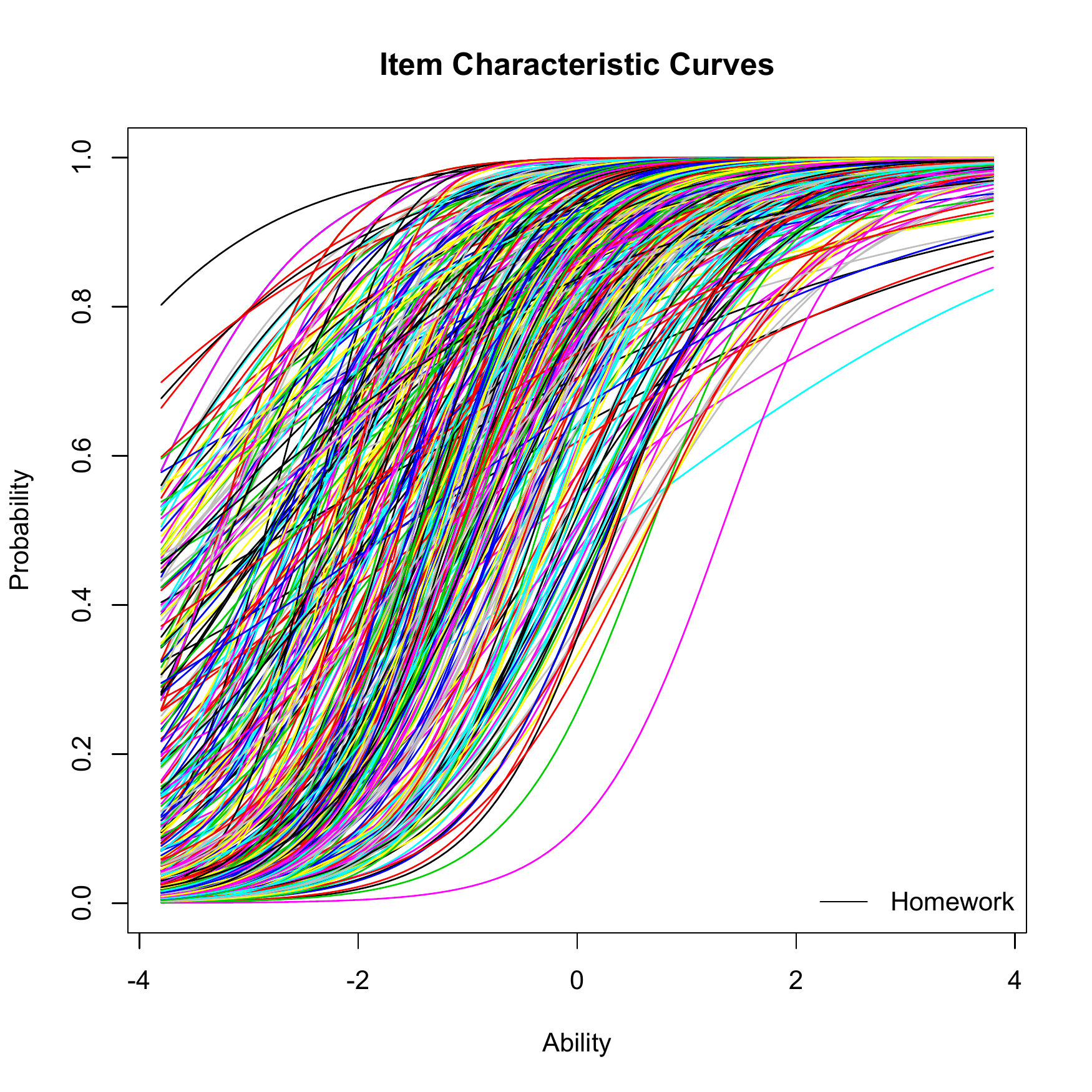}
\end{center}
\caption{Item characteristic curves for exam items (left panel) and homework items (right panel). Homework items were considered correct if the learner solved them eventually, regardless of number of attempts.}
\label{fig:cleancompare}
\end{figure*}

Fig.~\ref{fig:cleancompare} illustrates the difference in IRT results between controlled exam settings (IRT's natural habitat, left panel) and online homework settings (right panel), Each curve represents one problem and shows the likelihood of getting the problem correct as a function of the ability of the student. Both measures of difficulty and discrimination can be extracted from these typical ``Item Characteristic Curves:'' the further to the left the curve goes from low to high probability, the easier the problem, and the steeper the curve, the more strongly the item discriminates between low and high ability learners.

The exam outcome is rather typical: there are obviously some better and worse problems (most notably one problem that only half of the best students in the class are likely going to solve), but overall, the vast majority of the items exhibit the typical stretched S-shape. The homework outcome on the other hand is widely distributed. In particular, the high density of curves in the upper left quadrant of the figure shows that even low ability students get many problems correct.

\begin{figure}
\begin{center}
\includegraphics[width=0.45\textwidth]{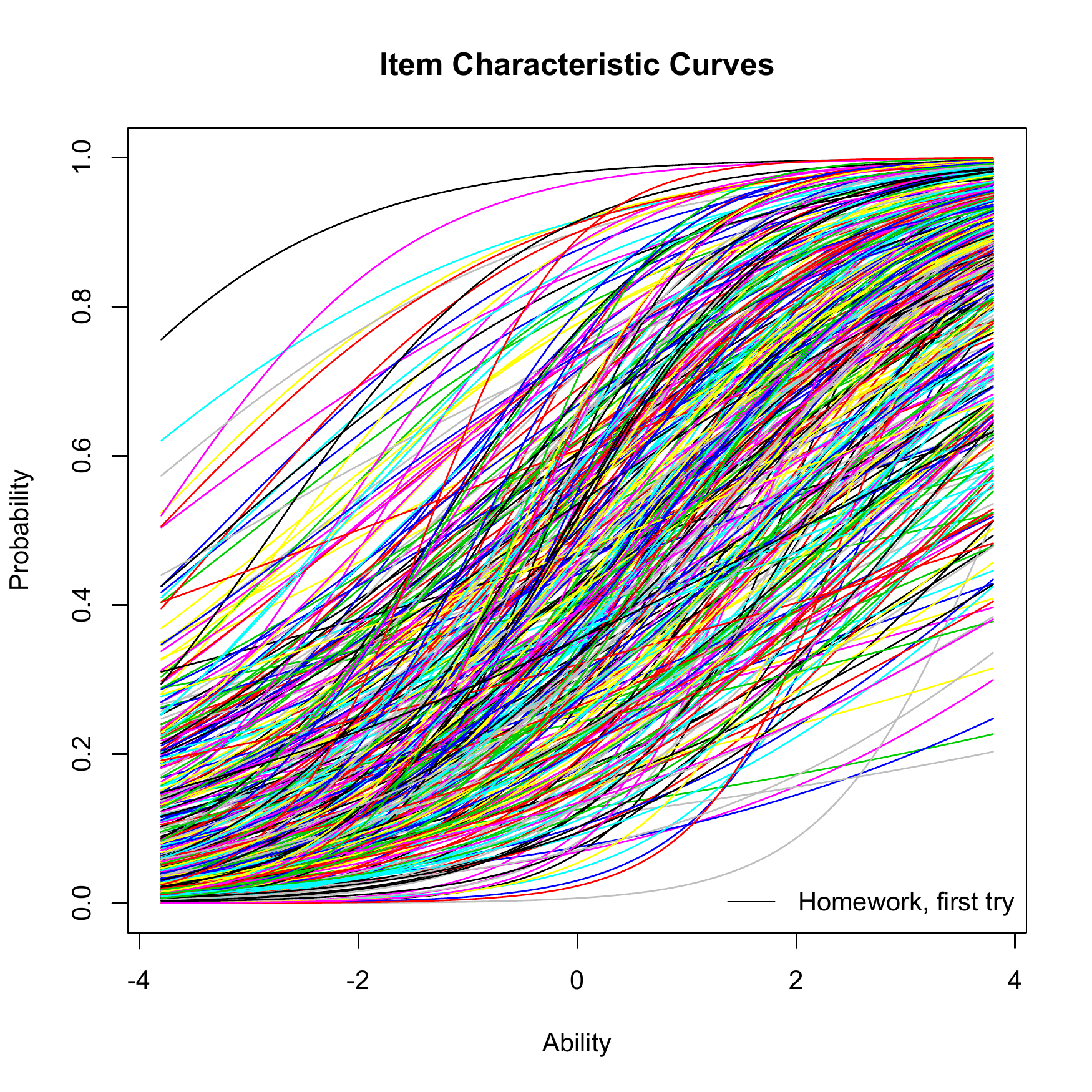}
\end{center}
\caption{Item characteristic curves for homework. Homework items were considered correct if they were solved on the first attempt.}
\label{fig:firsttry}
\end{figure}

It would be very easy to attribute this simply to the multiple allowed attempts, and in fact IRT studies of online homework tended to attempt to compensate by only considering the first attempt (e.g.~\cite{bergner12}), i.e., only considering the problem correct if it was solved on the first try. However, as Fig.~\ref{fig:firsttry} shows, using this approach, the curves spread out further, making the result even less meaningful.

While exam and homework problems are similar in character (except that all exam questions are multiple-choice), there is little direct overlap, as the exact same problem rarely appears on both homework and exams. Thus, unfortunately, in order to move beyond the qualitative comparisons of Figs.~\ref{fig:cleancompare} and~\ref{fig:firsttry}, a paired comparison between the IRT parameters is not possible.

\subsection{Learner Ability}\label{subsec:problemabi}
Without considering the larger question whether or not multiple-choice exams are a valid measure of learning, for the purposes of this study, the learner ability estimates based on the exam data will be considered a reliable measure. After all, these exams will largely determine the grade in the course, and the closer the estimates obtained from homework agree with the exam-based estimates, the more useful they are for early detection of students-at-risk.

The data allows for pair-wise comparison of ability score estimates, as the same students did work on homework and exams. With $i$ being the student, and with $A_{\mbox{\scriptsize E},i}$ and $A_{\mbox{\scriptsize H},i}$ being the ability estimates of the student for exams and homework, respectively, a measure for the overall similarity of the estimates is the cosine of the angle between the ``ability vectors,'' i.e., $\cos\theta=\vec{A}_{\mbox{\scriptsize E}} \vec{A}_{\mbox{\scriptsize H}} / (|\vec{A}_{\mbox{\scriptsize E}}||\vec{A}_{\mbox{\scriptsize H}}|)$ (a value of one would indicate perfect agreement, a value of zero would mean the outcomes are ``perpendicular'' (meaning, non-related), and a value of negative one would mean the exact opposite relationship). The correspondence turns out to be not very large: it is 0.64 between exams and homework considering all attempts, and 0.67 when only considering the first attempt.

\section{Influence of Multiple Attempts}\label{sec:attempts}
\subsection{Thresholds for Numbers of Attempts}\label{subsec:varytries}
In order to better understand the influence of imposing a threshold $N$ for the maximum number of considered attempts, different thresholds were employed. In other words, a problem was only considered correct if the number of attempts that the student needed to solve the problem was smaller than or equal to the respective threshold $N$. As it turns out, the 2PL-fit did not converge for several maximum numbers of attempts (namely, $N$ equals to 4, 8, 9, 10, 11, 12, and 13) , and was unstable for others (namely $N$ equals 3, 6, and 7). 
\subsubsection{Log-Likelihood}
\begin{figure}
\begin{center}
\includegraphics[width=0.45\textwidth]{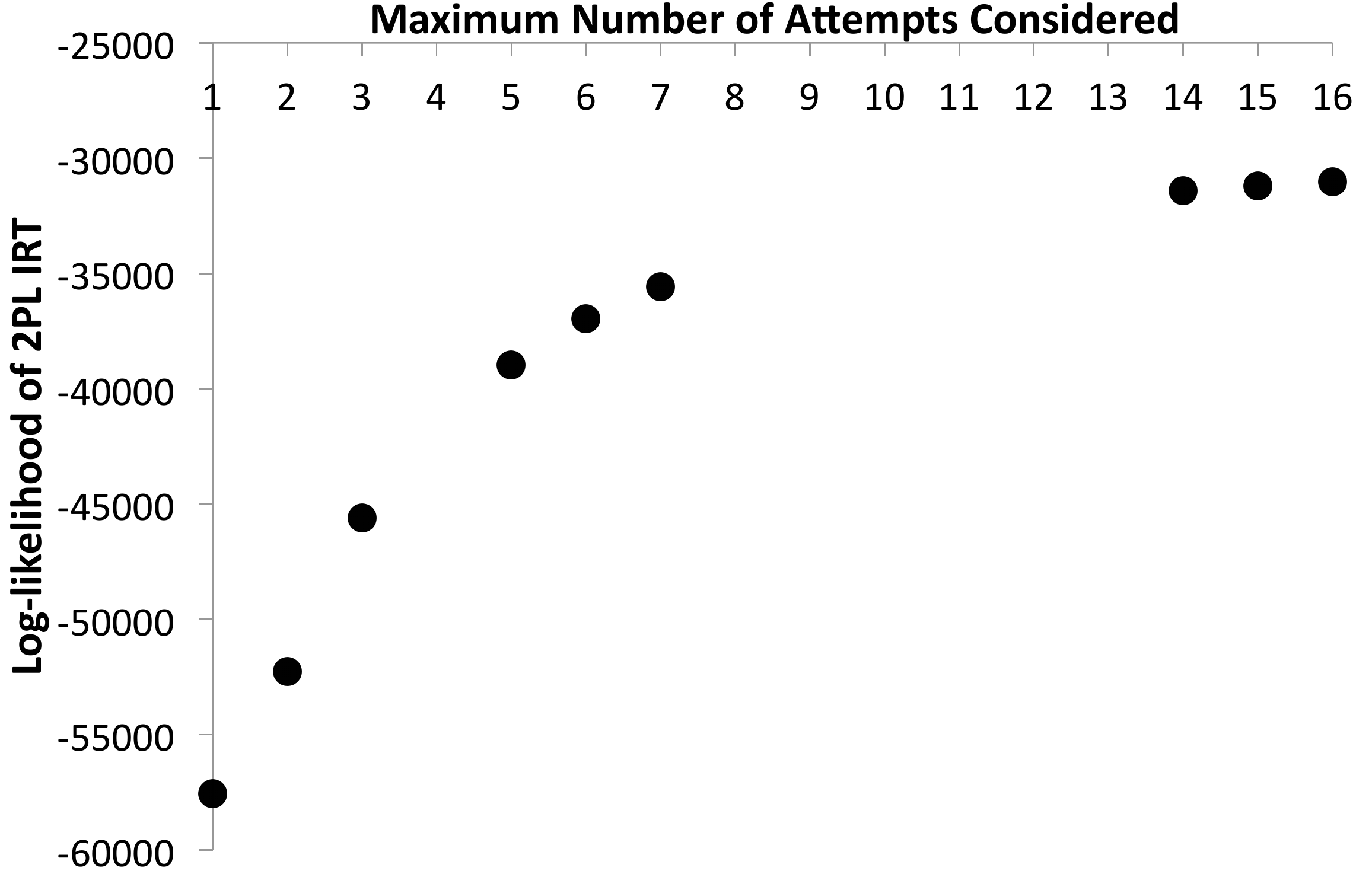}
\end{center}
\caption{Log-likelihood of the 2PL IRT analysis of the homework data as a function of the maximum number $N$ of considered attempts. Missing data points indicate that the fit did not converge. For comparison, without imposing a threshold on the number of attempts, the log-likelihood is -30070.}
\label{fig:loglikattempts}
\end{figure}

The log-likelihood of an IRT-fit can roughly be understood as a measure of the quality of the fit. Fig.~\ref{fig:loglikattempts} shows this value as a function of the imposed threshold. This value increases (i.e., the fit becomes more reliable) the more attempts are considered, and it eventually approaches the log-likelihood of the threshold-free analysis, i.e., -30070 (for comparison, the log-likelihood of the exam data  2PL-fit is -20930). 

\subsubsection{Item Difficulty}
\begin{figure}
\begin{center}
\includegraphics[width=0.45\textwidth]{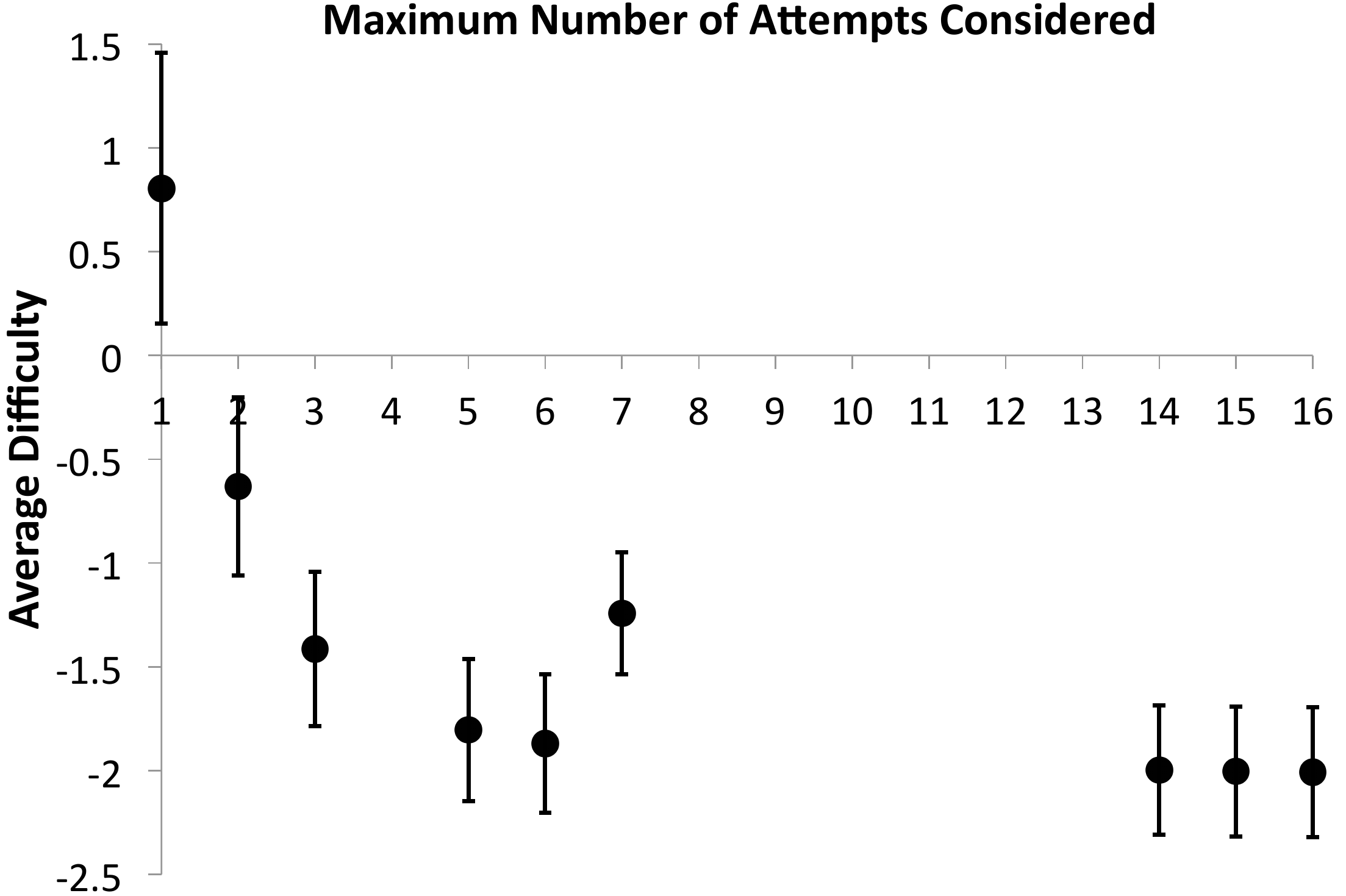}
\end{center}
\caption{Average difficulty of the homework items as a function of the maximum number $N$ of considered attempts. Missing data points indicate that the fit did not converge. The error bars indicate the average of the standard errors on the difficulties.}
\label{fig:difftries}
\end{figure}

Fig.~\ref{fig:difftries} shows the average difficulty of the items as a function of the maximum number $N$ of considered attempts. The 2PL-fits yield an error estimate of the item difficulties, and the error bars in the figure indicate the average of these error estimates. As expected, the problems get ``easier'' the more attempts are considered (the exception, $N$=7, is one of the unstable solutions). 

\subsubsection{Item Discrimination}\label{subsubsec:disctries}
\begin{figure}
\begin{center}
\includegraphics[width=0.45\textwidth]{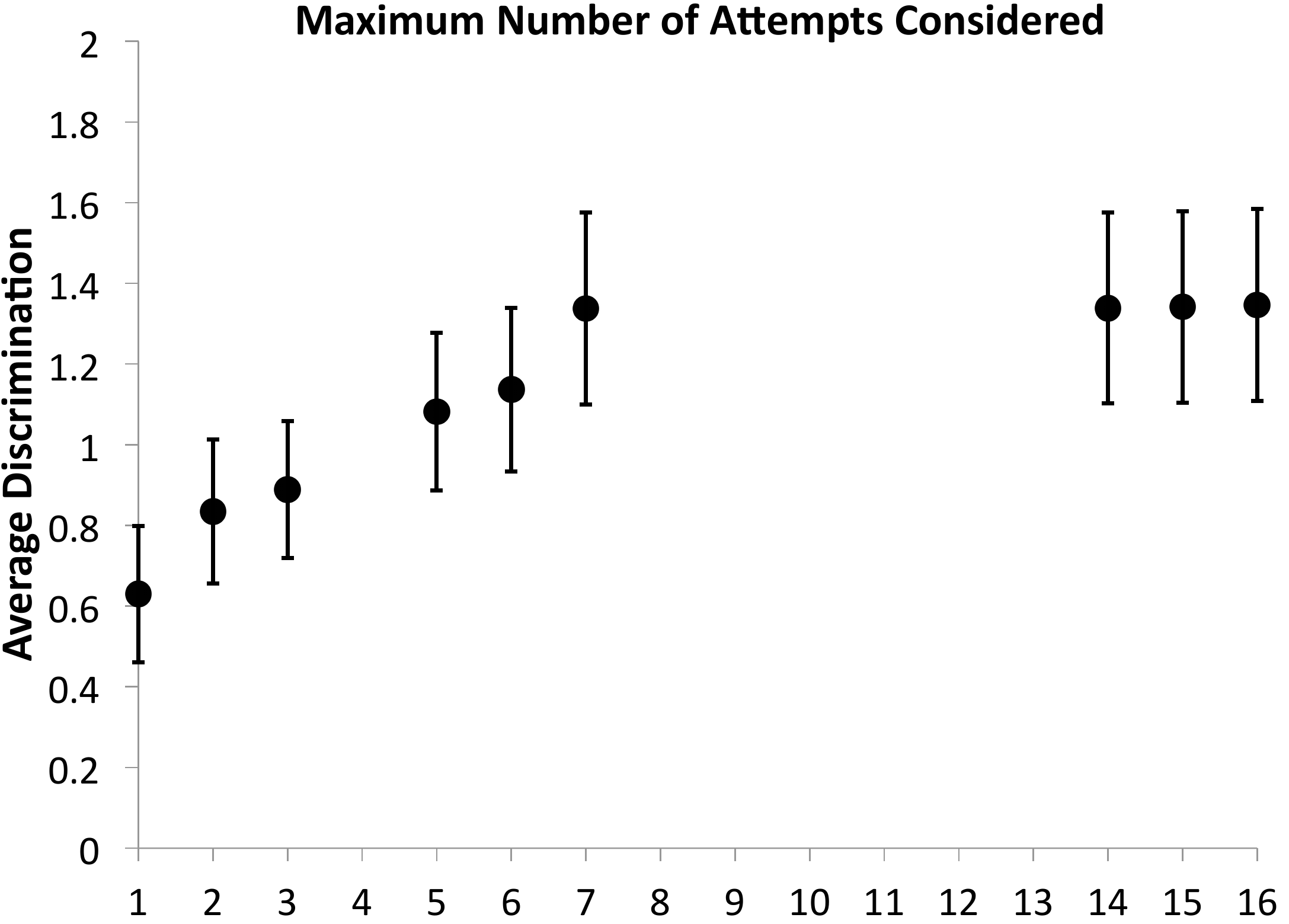}
\end{center}
\caption{Average discrimination of the homework items as a function of the maximum number $N$ of considered attempts. Missing data points indicate that the fit did not converge. The error bars indicate the average of the standard errors on the discriminations.}
\label{fig:disctries}
\end{figure}

Fig.~\ref{fig:disctries} shows the average discrimination of the items as a function of the maximum number of considered attempts, once again the error bars indicating the average of the error estimates. The result is interesting:  one may have expected the discrimination of the items would go down the more attempts are considered. In other words, a simple assumption would be that low and high ability students would best get separated by looking at the first attempt only (``good students would get the problem correct on the first try''), or maybe the second try (as even good students might be sloppy in their initial attempt). However, this is clearly not the case, which underlines the formative assessment character of online homework. If students indeed were expected to get everything right on the first attempt, there would be no need to assign the problem. It is clear that the discrimination actually increases as more and more attempts are considered, which may suggest that good students are distinguished by their tenacity to keep working on a homework problem, rather than their immediate genius.

\subsubsection{Learner Ability}\label{subsubsec:learnerabilitytries}
The cosine similarity between the ability estimates of the exam and homework data is essentially independent of the number of considered attempts, it varies between 0.67 for $N$=1 and 0.66 for $N$=16. Thus, the choice of $N$ has almost no influence on the quality of the estimate of the learner ability. For comparison, when considering all attempts, the cosine similarity was 0.64 (see Sect.~\ref{subsec:problemabi}), and thus not much additional information is gained by limiting the number of considered tries.

\subsection{Partial Credit for Multiple Attempts}\label{subsec:partialcredit}
Another approach is to allocate different amounts of partial credit for a solved problem, depending on how many attempts were needed to solve it. In this case, IRT needs to be extended to accept item results other than just zero and one, and one such model is the Generalized Partial Credit Model (GPCM)~\cite{gpcm}, which is also implemented in the {\tt ltm}-package. The model only deals with polytomous data, so continuous mappings of numbers of attempts $n$ to partial credit $p$ (e.g., $p=1/n$ or $p=\exp(1-n)$) are not supported. To explore this mechanism, a simple mapping of $p=6$ for having solved the problem on the first try, $p=5$ for the second try, etc, to $p=2$ for successful completion after any number of tries was chosen. For technical reasons, $p=1$ was used for unsuccessful attempts, as the package translates $p=0$ into ``not answered.''

The resulting fit had a low log-likelyhood (-126,370). Item discriminations were reported on a different scale than in the dichotomous model, but the cosine similarity is very high: it is 0.94 for considering either all attempts or up to $N$=16 attempts (see Sect.~\ref{subsubsec:disctries}), and 0.95 for considering only the first attempt ($N$=1); regarding the item discrimination, the dichotomous and polytomous models are thus almost equivalent. Regarding learner abilities, the fit yielded a cosine similarity of 0.66 with the exam data, which is almost exactly equivalent to considering all attempts or the findings in Sect.~\ref{subsubsec:learnerabilitytries}  --- once again, both models are almost equivalent.

For our data (256 students, 401 items, and 6 partial credit bins), computation time increased by almost two orders of magnitude compared to the simple dichotomous model without gaining additional information about the items or the learners. Thus, the additional computational complexity does not appear to be justified compared to simply considering the dichotomous model whether or not the learner eventually solved the problem.

\subsection{Discussion of Multiple Attempts}
Essentially all attempts of remedying the effect of offering multiple attempts by limiting the number of considered attempts or employing some partial credit scheme were un- or even counter-productive. Considering only the first attempt did not lead to any improvements, which is likely due to the typical ``guessing'' behavior of students which leads to essentially wasting the first attempt half of the time~\cite{kortemeyer09} (and, 25-35\% on later tries~\cite{laverty12b})  --- the students know that they have multiple attempts and act accordingly. Beyond considering only the first attempt, the agreement with exam data regarding problem discrimination and learner ability is equivalent for practical considerations, and essentially nothing is gained beyond simply looking at the dichotomous model of whether or not the students eventually solved the problem.

The good news is that this robustness of results proves the value of using IRT in the online homework context, since the results of simple descriptive statistics would vastly depend on these choices: for example, as more and more attempts are considered, the classical descriptive values of the abilities of the learners increase --- a non-sensical situation that would demand somewhat arbitrary choices of  thresholds or partial credit schemes. 

\section{Guessing and Copying}\label{sec:guessing}
\subsection{Three Parameter Model}
When estimating the effect of guessing, a natural first approach is to employ a three parameter model (TPM), where the third parameter models the guessing behavior~\cite{birnbaum}. As it turns out, though, this third parameter only leads to minimal improvement of the fit, regardless of whether only the first or all attempts are considered. When considering only the first attempt, some of the problem discrimination gets lost in favor of the guessing parameter, but this effect almost completely vanishes when considering all attempts. In both cases, the average guessing parameter is small (0.031 when considering only the first attempt, and 0.019 when considering all attempts). The cosine similarity of the learner ability with the exam results once again remains largely unchanged.

Initially, particularly when only considering the first attempt, this outcome seems counter-intuitive --- do students not guess after all? However, the guessing modeled by the three parameter model is non-random guessing in the case of difficult problems. This is not what students are doing: their first attempt is frequently truly random or at most reflecting some ``gut feeling,'' regardless of the actual difficulty of the problem~\cite{kortemeyer09}. In addition, IRT expects a limited number of answer choices, where the hit-and-miss ratio is determined by the number of bubbles, not a free-response with essentially infinitely many answer choices. In summary, a three parameter model with a guessing parameter is hardly useful in the online homework scenario and in fact might be overfitting the data.

\subsection{Earlier Parts of the Semester}\label{subsec:earlypart}
Students often begin a new course with good intentions and high scores in the diligence cluster of epistemological surveys, only to have these good intentions fall by the wayside as the semester progresses~\cite{mpex}. It is thus reasonable to assume that guessing and copying behavior increases over the course of the semester, which was indeed found to be the case~\cite{palazzo10}.

To investigate if a signature of this can be found in the homework data, only the first quarter of the semester was considered. As it turns out, the cosine similarity of the learner abilities to the exam data decreases when considering all attempts (0.60), but it increases when only considering the first attempt (0.72). In other words, in the first weeks of the semester, the first attempt on a homework problem is still meaningful --- in fact, the extracted cosine similarity of the learner ability to the exam data is the highest found in this study. As the semester progresses, it appears that  considering all attempts increases the quality of the learner ability estimates (to 0.64, see Sect.~\ref{subsubsec:learnerabilitytries}), while considering only the first attempt becomes less useful (the cosine similarity decreases to 0.67). These effects are minimal, and the absolute agreement is moderate at best in any case, but they do support the emerging picture of the effect of copying and guessing.

The good news is that an early-warning system for learners-at-risk might indeed be possible. Using IRT, already a few weeks into the semester, meaningful learner ability estimates can be obtained (in this case, considering only the first attempt does improve the quality of the estimate).

\subsection{Modeling Guessing and Copying}
While the effect of offering multiple attempts can to some degree be ``undone'' in the homework data by either limiting the number of considered attempts (Sect.~\ref{subsec:varytries}) or employing a partial credit model (Sec.~\ref{subsec:partialcredit}), guessing and copying cannot be subtracted from the homework data beyond limiting the analysis to the initial weeks of the course where the students still ``behaved better'' (Sect.~\ref{subsec:earlypart}). Instead, to assess their effect, they can be artificially added to the exam data; assuming that the exam data is ``clean,'' guessing and copying behavior can be simulated by manipulating the student/item response matrix.

In an exam scenario, copying would mean copying somebody else's answers without necessarily knowing if these answers are right or wrong. The behavior would need to be modeled by increasingly copying data from one row of the student/item-matrix into another row, i.e., increased copying would make one student's item response row increasingly similar to the row of another student. Copying of online homework with immediate feedback is different: a student would not copy somebody else's incorrect answer. Thus, in the online homework scenario, guessing and copying are quite similar, only the success rate is much higher for the latter.

Thus, both guessing and copying of online homework with immediate correctness feedback can be modeled approximately with the same two-parameter manipulation of the ``clean'' exam data. The two parameters are percentage of occurrence, i.e., how many percent of the problems the student is guessing on or copying, and success rate, i.e., how often this behavior yields a correct answer. The success rate ranges from low percentages for ``stabs in the dark'' and somewhat higher percentages for ``educated guesses'' to very high percentages for sure-fire solutions, i.e., solutions that have been copied or adapted from somebody else's correct solution (note that the problems are randomized, so the success rate for copying is likely not 100\%).

\begin{figure}
\begin{center}
\includegraphics[width=0.45\textwidth]{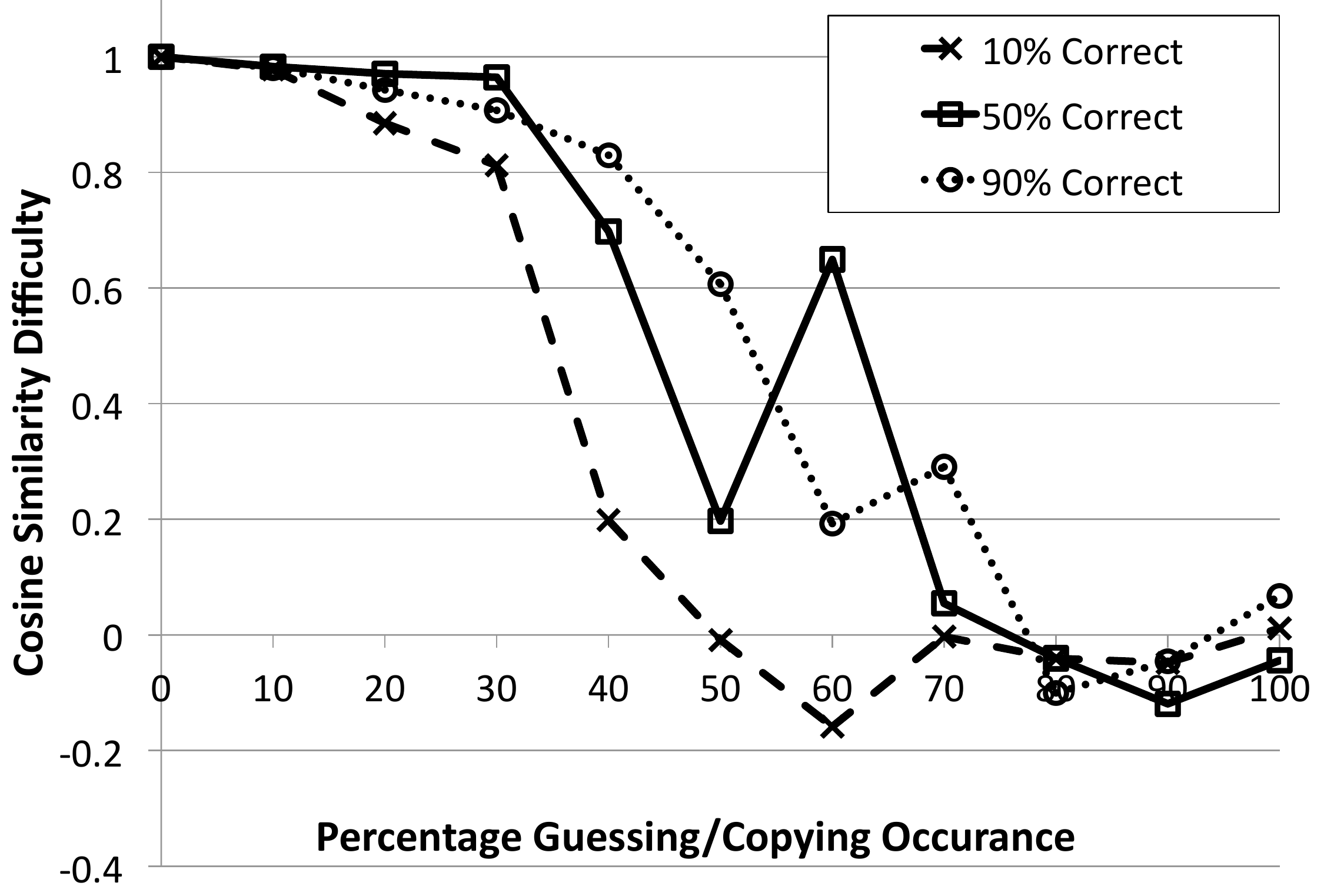}
\end{center}
\caption{Cosine similarity of the item difficulties from the ``clean'' exam values as a function of simulated copy/guess occurrence. Values are shown for a success rate of 10\% (``educated guessing''), 50\%, and 90\% (``copying'').}
\label{fig:guessdiff}
\end{figure}

\begin{figure}
\begin{center}
\includegraphics[width=0.45\textwidth]{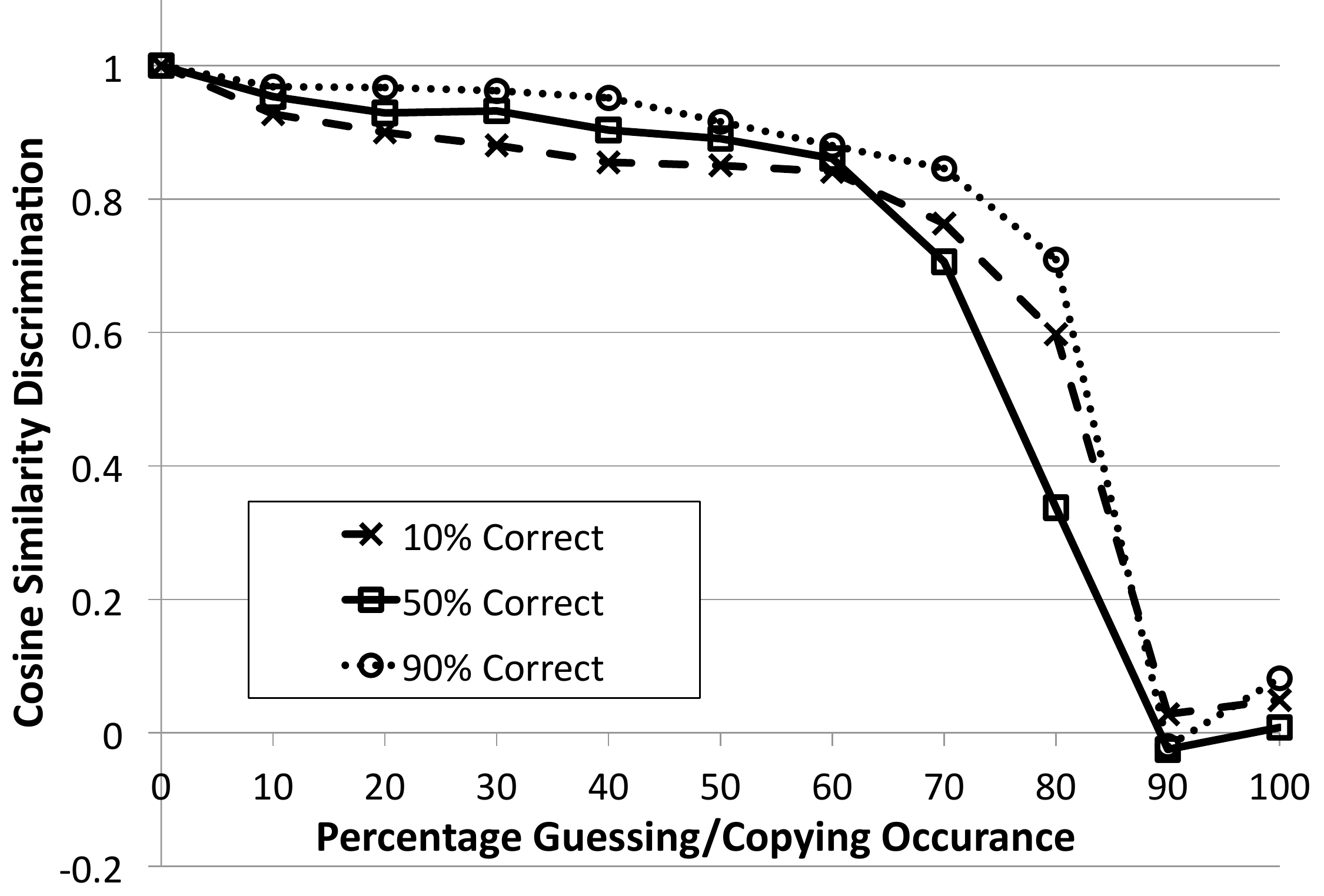}
\end{center}
\caption{Cosine similarity of the item discriminations from the ``clean'' exam values as a function of simulated copy/guess occurrence. Values are shown for a success rate of 10\% (``educated guessing''), 50\%, and 90\% (``copying'').}
\label{fig:guessdisc}
\end{figure}

\begin{figure}
\begin{center}
\includegraphics[width=0.45\textwidth]{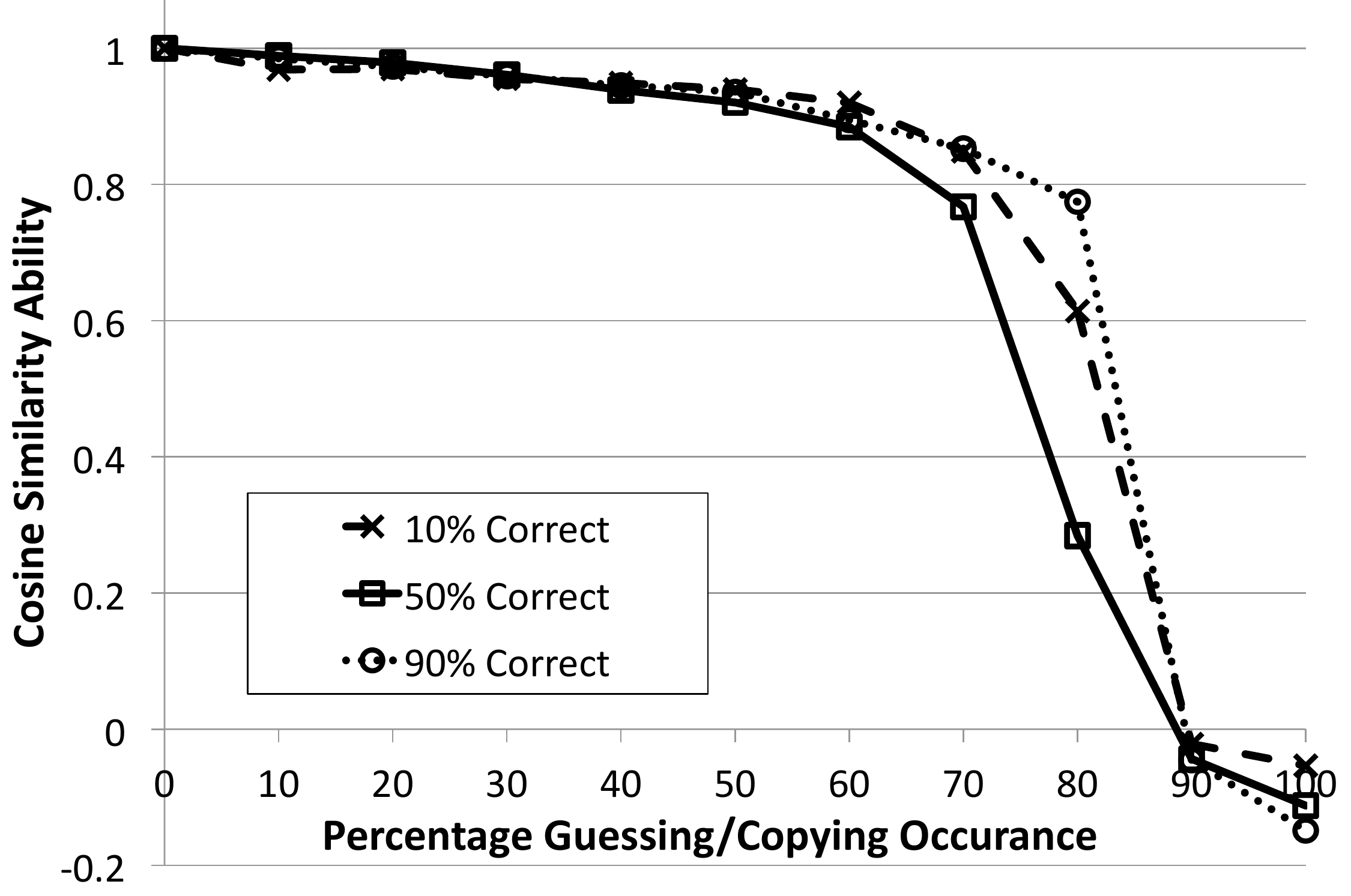}
\end{center}
\caption{Cosine similarity of the learner abilities from the ``clean'' exam values as a function of simulated copy/guess occurrence. Values are shown for a success rate of 10\% (``educated guessing''), 50\%, and 90\% (``copying'').}
\label{fig:guessabi}
\end{figure}

Using a 10\% success rate for educated guesses, a 90\% success rate for copying randomizing homework, and a 50\% success rate for comparison, the decay of the ``clean'' estimates from the original exam data was traced for different occurrence rates between 0\% and 100\%. Figure~\ref{fig:guessdiff} shows the effect on the problem difficulty estimates. Already if more than a third of the problem solutions are guessed or copied, the cosine similarity of the difficulty estimates rapidly decreases. In other words, problem difficulties are very sensitive to the introduced noise.

On the other hand, Fig.~\ref{fig:guessdisc} and Fig.~\ref{fig:guessabi} show good agreement with the original discrimination and ability estimates until more then two thirds of the problem solutions are guessed or copied. Going back to the results from the real homework data, where cosine agreements of learner ability with exam data are in the range of 0.60 to 0.72, one might be tempted to conclude that guessing and copying are more prevalent than assumed thus far --- however, that is not taking into account that the homework and exam problems are different, as well as the other confounding effects of multiple attempts and free-response solutions.

Once again, the good news is the robustness of IRT estimates in the presence of noise. While the difficulty estimates might break down somewhat early, the discrimination and  learner ability estimates show remarkable reliability.

A more sophisticated model should probably assume that some students guess/copy and others do not (instead of the indiscriminate guessing/copying in the current model), or that maybe low-ability students are more prone to guess or that they would likely copy from high-ability students, but these might better be left to a later study. The main result from this simple model is that IRT parameters and ability estimates are relatively robust with respect to the noise introduced by copying and guessing over varying ranges, but then very rapidly break down --- there is no graceful degradation.

\section{Conclusions}\label{sec:conclusions}
Extending IRT to online homework scenarios poses considerable challenges. It should not simply be assumed that the model transfers, and this study attempted to quantify the effects of violating the original assumptions of the model.

When using IRT to estimate learner ability, results obtained from online homework show the correct trends compared to exam data, but the absolute agreement is moderate. Within these limitations, however, IRT is remarkably robust with respect to online homework offering multiple attempts, and stable in terms of noise introduced from copying and guessing up to a point. When it comes to item parameters, discrimination is less affected by the homework environment than difficulty; however, when IRT breaks down, it does so rapidly.

The agreement of homework and exam data, as well as the quality of item parameters, is not improved by using more complicated models or filtering of the data. In particular, trying to model guessing or introducing partial credit does not yield additional information. The effect of only considering the first attempt on homework wears out over the course of the semester. Thus, overall it seems that simple is better: using a straightforward two parameter logistic model on the full homework (only considering if the problems were solved eventually) appears to be about as fruitful as the introduction of additional assumptions or constraints.

\bibliographystyle{apsper}
\bibliography{bibfile}

\end{document}